\documentclass{ieeeaccess}
\usepackage{cite}
\usepackage{amsmath,amssymb,amsfonts}
\usepackage{algorithmic}
\usepackage{graphicx}
\usepackage{array}
\usepackage[hidelinks,bookmarks=false]{hyperref}
\newcommand{\orcidlink}[1]{\href{https://orcid.org/#1}{\includegraphics[height=1.5ex]{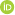}}}
\usepackage{textcomp}
\usepackage{url}

\usepackage{xcolor}
\newif\ifshowrev
\showrevfalse          
\ifshowrev
  \newcommand{\rev}[1]{\textcolor{red}{#1}}           
  \newenvironment{revblock}{\color{red}}{}             
\else
  \newcommand{\rev}[1]{#1}
  
\fi

\usepackage{bm}

\makeatletter
\def\bstctlcite{\@ifnextchar[{\@bstctlcite}{\@bstctlcite[@auxout]}}
\def\@bstctlcite[#1]#2{\@bsphack\@for\@citeb:=#2\do{\edef\@citeb{\expandafter\@firstofone\@citeb}\if@filesw\immediate\write\csname #1\endcsname{\string\citation{\@citeb}}\fi}\@esphack}
\makeatother
\makeatletter
\AtBeginDocument{\DeclareMathVersion{bold}
\SetSymbolFont{operators}{bold}{T1}{times}{b}{n}
\SetSymbolFont{NewLetters}{bold}{T1}{times}{b}{it}
\SetMathAlphabet{\mathrm}{bold}{T1}{times}{b}{n}
\SetMathAlphabet{\mathit}{bold}{T1}{times}{b}{it}
\SetMathAlphabet{\mathbf}{bold}{T1}{times}{b}{n}
\SetMathAlphabet{\mathtt}{bold}{OT1}{pcr}{b}{n}
\SetSymbolFont{symbols}{bold}{OMS}{cmsy}{b}{n}
\renewcommand\boldmath{\@nomath\boldmath\mathversion{bold}}}
\makeatother

\begin{document}
\bstctlcite{IEEEexample:BSTcontrol}
\history{Received 22 September 2026, accepted 11 October 2026, date of publication 15 October 2026, date of current version 28 October 2026..}
\doi{10.1109/ACCESS.2026.3743834}

\title{Automatic Knowledge Graph Construction and Query for Earthquake Catalogs}

\author{\uppercase{Yuxin Zhou}\,\orcidlink{0009-0000-2877-9173}\authorrefmark{1,2},
\uppercase{Huai Zhang}\,\orcidlink{0000-0003-0411-4841}\authorrefmark{1},
AND \uppercase{S. Mostafa Mousavi}\,\orcidlink{0000-0001-5091-5370}\authorrefmark{2}}

\address[1]{State Key Laboratory of Earth System Numerical Modeling and Application, Department of Earth and Planetary Sciences, University of Chinese Academy of Sciences, Beijing 100049, China (e-mail: hzhang@ucas.ac.cn)}
\address[2]{Department of Earth and Planetary Sciences, Harvard University, Cambridge, MA 02138 USA (e-mail: yzhou2@fas.harvard.edu)}
\tfootnote{This work was supported in part by the Harvard Milton Fund.}

\markboth
{Zhou \headeretal: Automatic Knowledge Graph Construction and Query for Earthquake Catalogs}
{Zhou \headeretal: Automatic Knowledge Graph Construction and Query for Earthquake Catalogs}

\corresp{Corresponding author: Huai Zhang (e-mail: hzhang@ucas.ac.cn).}

\begin{abstract}
In recent years, the number of events in earthquake catalogs has significantly increased due to the utilization of more effective \rev{deep-learning-based detectors and phase pickers, but answering} open-ended questions such as ``what characterizes this sequence?'' remains constrained by rigid spatiotemporal windowing and subjective expert interpretation. \rev{We apply graph-based retrieval-augmented generation (GraphRAG) directly to raw, tabular catalog records across three independent catalogs}: a reservoir-adjacent swarm (Qiaojia-Dongchuan), the 2019 Ridgecrest tectonic sequence, and the 2021 Maduo $M_\mathrm{w}$~7.4 aftershock sequence, to build structurally complete, queryable knowledge graphs without the need for manual data structuring. \rev{We scored 1,500 answers against ground truth computed from the catalogs and audited the extracted graphs record by record against the catalogs and a rule-based reference graph. This evaluation exposes several failure modes; seismology-informed prompt fixes remove the targeted fabrication modes and raise mechanism-question scores to 2.85/3 at best.} GraphRAG thus provides a practical, low-cost interface for querying raw earthquake catalogs, with a construction pipeline that transfers readily across catalogs and seismology-specific prompting that improves response reliability, although persistent limitations remain. \rev{By systematically identifying both the capabilities and failure modes of a widely used large language model (LLM)-based tool, this study provides the seismological community with practical guidance while highlighting the limitations that must be considered when adopting emerging LLM tools for seismicity analysis.}
\end{abstract}

\begin{keywords}
Earthquake catalog, GraphRAG, knowledge graph, large language model, retrieval-augmented generation.
\end{keywords}

\titlepgskip=-21pt

\maketitle

\section{Introduction}
\label{sec:introduction}
\PARstart{D}{ense-array} monitoring \cite{ross_searching_2019,shelly_highresolution_2020} and deep-learning phase detection/picking \cite{zhu_phasenet_2019,mousavi_earthquake_2020,Ross_2018,mousavi_deep-learning_2022} have together driven an order-of-magnitude expansion in earthquake-catalog size; for example, the 2021 Maduo $M_\mathrm{w}$~7.4 aftershock sequence contains over 10,000 relocated events \cite{Guan_2024}. Beyond scale, a deeper challenge is characterizing what high-resolution catalogs represent physically: mainshock-aftershock decay \cite{gutenberg_frequency_1944,Utsu_1995}, swarms, foreshock sequences identifiable only in hindsight, and induced seismicity \cite{gupta_review_2002} are still distinguished mainly by coarse qualitative criteria, or by point-process models fit after the fact \cite{Ogata_1988}, which do not allow quantitative, reproducible cross-sequence comparison in real time. However, this increase in the typical \rev{sizes} of earthquake catalogs has not yet helped with this major challenge. Our analyses and interpretations of the high-resolution deep-learning-based catalogs remain largely limited to a few classical engineered features.

Knowledge graphs organize semantically linked information as reasoning-ready (entity, relationship, entity) triples \cite{Hogan_2021}, and have been applied to earthquake emergency response \cite{qiu_earthquake_2024} and seismic metadata organization \cite{davis_knowledge_2024}. Retrieval-augmented generation (RAG) \cite{Lewis_2020} is an artificial intelligence architecture that \rev{aims to improve the factual accuracy of} large language models (LLMs) by grounding their responses in dynamically retrieved, domain-specific external data. Microsoft's GraphRAG \cite{edge_local_2025} automatically extracts entities and relationships, detects communities via the Leiden algorithm \cite{Traag_2019}, and generates natural-language summaries and query access, with no predefined ontology\rev{, as part of a broader effort to unify LLMs and knowledge graphs} \cite{Pan_2024}. This raises a natural question: can GraphRAG be applied directly, end-to-end, to earthquake catalogs of varying scale, region, and character to produce a working natural-language query interface, and can its most damaging failure modes be suppressed with modest, seismology-specific engineering effort rather than a full custom rebuild?

RAG has recently begun to be applied within seismology and geohazard research specifically. Yao \emph{et al.} \cite{yao_knowledge_2026} combine knowledge-graph construction with a hybrid RAG strategy for earthquake emergency response, extracting entities and relationships from thousands of professional emergency-management documents; Polo-Bautista \emph{et al.} \cite{polo-bautista_harnessing_2025} use LLMs to build knowledge graphs from earthquake news articles. Recent reviews \cite{yu_rag_2025,li_toward_2026} note that RAG in the geosciences remains applied mainly to narrative, prose-style sources (news, reports, technical literature) rather than to the raw, tabular observational records a catalog itself consists of, and call for RAG architectures tailored to structured geoscientific data. Our study differs in three respects: it applies RAG directly to structured, numerical catalog records rather than narrative text describing events after the fact; it targets fully automatic, schema-free construction with no per-catalog ontology or extraction-rule engineering, as opposed to the hand-designed ontologies underlying prior earthquake knowledge-graph work; and it evaluates GraphRAG specifically, whose automatic community detection and hierarchical summarization are built to transform thousands of brief catalog records into sequence-level summaries, \rev{a capability our benchmark evaluates directly.}

Section~\ref{sec:capability} demonstrates that fully automatic indexing and natural-language querying run end-to-end on all three catalogs under default prompts, establishing mechanical transferability while exposing low baseline answer quality \rev{(catalog-only averages 0.78--1.38/3)}. Section~\ref{sec:enrichment} quantifies the catalog-dependent effect of narrative-text enrichment. Section~\ref{sec:prompt} describes an iterative, seismology-oriented prompt-engineering scheme that eliminates the targeted fabrication modes across all 600 post-fix answers. \rev{That section also introduces a baseline vector-RAG comparison} (embedding-similarity retrieval over the same text chunks, with no graph layer) \rev{to quantify the community-report layer's effect on catalog-wide questions.} Section~\ref{sec:residual} details the specific catalog errors \rev{identified during scoring, such as the systematic misdating} of the Ridgecrest mainshock and the hallucinated inclusion of an unlisted Maduo event. We report these issues to guide safe deployment, rather than as the study's central finding. \rev{Extended examples, the per-question scores of every answer, and the prompt modifications are provided in the Supplementary Materials (SM).}

\section{Materials and Methods}
\label{sec:methods}

\subsection{Case Studies}

\begin{figure*}[t]
\centering
\includegraphics[width=.80\textwidth]{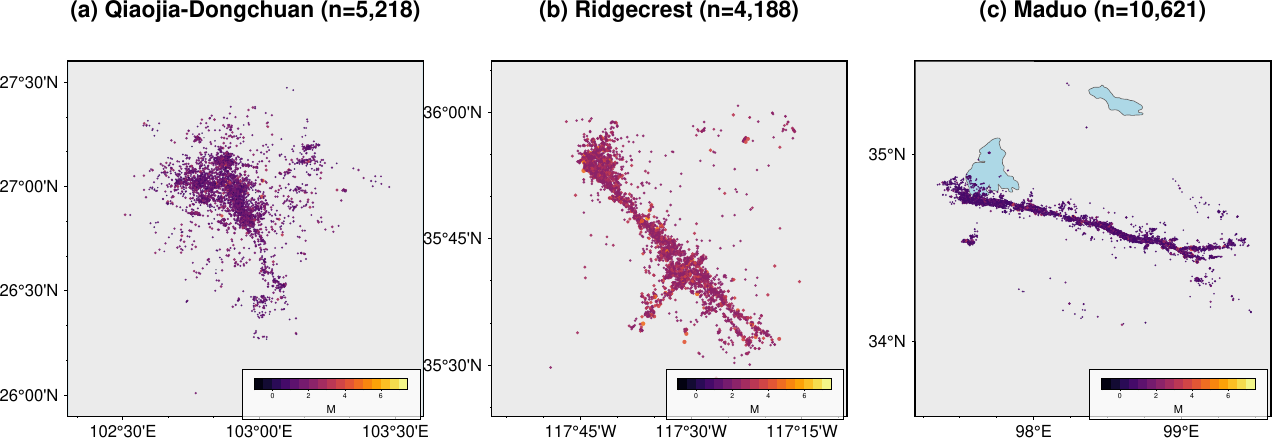}
\caption{\textbf{Spatial distribution of seismic events. (a) Qiaojia-Dongchuan; (b) Ridgecrest; (c) Maduo. Epicenters colored by magnitude.}\label{fig01}}
\end{figure*}

We use three independent catalogs differing in scale, region, and origin, totaling 20,027 events (Fig.~\ref{fig01}; overview table in SM Table~S1): (1) the Qiaojia-Dongchuan relocation catalog, 5,218 events recorded by a temporary dense array adjacent to the Baihetan Reservoir between 23~August~2022 and 17~March~2023; (2) the 2019 Ridgecrest sequence, 4,188 events ($M\!\geq\!2.0$) spanning July 2019, a standard catalog \cite{shelly_highresolution_2020} on a purely tectonic strike-slip system in the Mojave Desert; and (3) the 2021 Maduo $M_\mathrm{w}$~7.4 sequence, the largest of the three at 10,621 events ($M\!\geq\!0.5$, 1~June~2021--8~June~2023), relocated by Guan \emph{et al.} \cite{Guan_2024}. This aftershock-only catalog begins after the true 22~May~2021 mainshock and contains no May-2021 data.

\subsection{GraphRAG Pipeline and Evaluation Design}

\begin{figure*}[t]
\centering
\includegraphics[width=.87\textwidth]{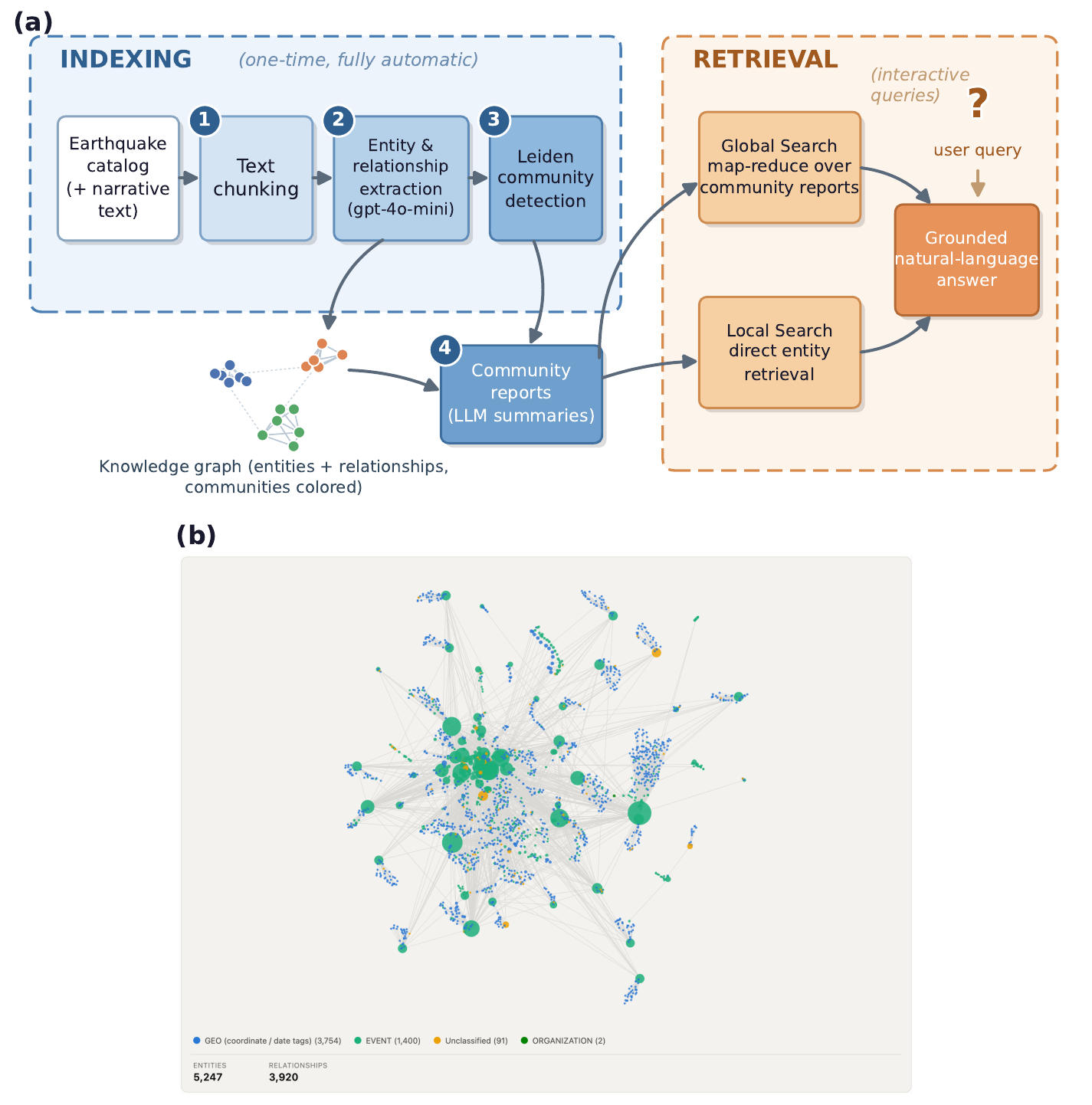}
\caption{\textbf{(a) Schematic of the GraphRAG indexing and retrieval pipeline. (b) Entity-relationship graph automatically constructed by GraphRAG from the Qiaojia catalog (5,247 entities; 3,920 relationships), with no hand-designed schema; GEO (coordinate/date tags), EVENT (individual earthquake records), and a small residual of Unclassified/Organization entities from generic extraction.}\label{fig02}}
\end{figure*}

The system architecture relies on two primary phases: Indexing and Retrieval (Fig.~\ref{fig02}a). The indexing phase turns the raw catalog rows into an entity-relationship graph. It begins with LLM-based entity and relationship extraction (using \texttt{gpt-4o-mini}) to identify seismic features such as individual earthquakes, fault structures, and their links in space and time. Next, Leiden community detection groups these interconnected events into communities. The system then processes these clusters through automated community-report generation, creating high-level textual summaries that describe the defining characteristics of each localized sequence.

During the retrieval phase, the system supports two
complementary search modes for interacting with the
indexed seismic data. Global Search \rev{synthesizes} across all community reports, making it \rev{suited to} broad, sequence-level questions regarding
overarching migration patterns or aggregate statistics. In
contrast, Local Search executes precise entity retrieval
to isolate exact details concerning specific mainshocks\rev{, individual records,} or localized catalog anomalies.

Evaluation proceeds in three stages: (i) capability demonstration under default prompts (Section~\ref{sec:capability}); (ii) narrative-text enrichment, evaluated with a 100-question benchmark spanning five categories\rev{:} (A) overall summarization, (B) precise statistics, (C) local retrieval, (D) physical mechanism, and (E) temporal-stage comparison; 20 questions each\rev{; the full question list is given in SM Table~S8} (Section~\ref{sec:enrichment}); and (iii) iterative, seismology-oriented prompt engineering targeting the exposed failure modes (Section~\ref{sec:prompt}). \rev{A rule-based knowledge graph built from the same Qiaojia-Dongchuan catalog (10,766 nodes; 113,620 edges; SM Text~S7) reproduces all 44 deterministic reference answers and serves as the reference for the graph-quality evaluation of Section~\ref{sec:capability} and as the deterministic-query comparator of Section~\ref{sec:conclusions}.} To provide a point of comparison, we identically scored a standard vector-RAG baseline. This baseline used the same text chunks, embeddings, and LLM as our GraphRAG pipeline, but relied on top-10 chunk similarity for retrieval rather than a graph layer.

\rev{All 1,500 answers (1,200 GraphRAG and 300 baseline responses) were scored from 0 to 3 by the same scoring procedure (SM Text~S6). Deterministic checks recompute every catalog statistic and flag out-of-range values, and an LLM judge (gpt-4.1, distinct from the answering model) grades each answer against these facts, the narrative text where applicable, and a per-question reference computed from the catalog. Two seismologists independently scored a blind, stratified sample of 150 answers (SM Text~S6.5).} Responses were graded using the following rubric: a score of 3 indicates a correct, independently verifiable central claim; a score of 2 denotes a correct central claim grounded in the data, but containing secondary errors, omissions, or unverifiable elements; a score of 1 applies when the central claim is wrong or missing (e.g., refusing an answerable question or misrepresenting data subsets) despite using genuine records; and a score of 0 is reserved for fabricated values, hallucinated records, or false claims presented as factual. Refusals earn a 3 only if the required information is completely absent from the catalog.

\rev{To quantify uncertainty, 95\% bootstrap confidence intervals were computed for every condition mean (SM Tables~S5--S7). The catalog-only gain of the full prompt-engineering scheme over the default prompts is Qiaojia +0.04, Ridgecrest +0.13, and Maduo +0.49, and the difference between GraphRAG and the vector-RAG baseline is Qiaojia $-0.05$, Ridgecrest +0.02, and Maduo $-0.07$: the fixes produce a gain that survives the paired test on one of the three catalogs, and GraphRAG and the baseline are statistically indistinguishable overall (SM Table~S7).}

\section{Results: Capability Demonstration}
\label{sec:capability}

With no hand-designed schema and no catalog-specific configuration, GraphRAG automatically built structurally complete, queryable knowledge graphs for all three catalogs from raw tabular records alone: Qiaojia-Dongchuan (5,247 entities, 3,920 relationships, 146 communities; Fig.~\ref{fig02}b), Ridgecrest (1,985 entities, 2,449 relationships, 83 communities), and Maduo (15,402 entities, 11,784 relationships, 600 communities). Entity types are consistently dominated by coordinate/date (GEO) tags ($\sim$72--83\%) across all three, confirming that indexing capability is general and does not depend on catalog region or scale. Per-catalog entity-type and community-size distributions are given in the SM (Figures S1--S2). \rev{Here, an ``entity'' is any distinct item extracted from the catalog text, not necessarily an individual earthquake. With the default prompts, only 0.1--16\% of records are represented as individual entities, whereas the seismology-specific prompts in Section~\ref{sec:prompt} correctly represent 97--99\% of records, with 99.7--99.9\% exact magnitude agreement and only 0.1--3.5\% duplicates. Comparison with a rule-based graph further shows that the resulting relations and communities primarily capture temporal rather than physical structure: 98--100\% of communities correspond to consecutive catalog windows, without comparable spatial clustering (SM Text S7; Table S13).} 

The resulting graphs support natural-language queries that a fixed-field database cannot: Local Search queries retrieve precise, traceable individual records (e.g., correctly naming Ridgecrest's $M7.1$ mainshock and its 6~July date), and Global Search synthesizes catalog-wide descriptions of sequence character. However, this out-of-the-box functionality has reliability issues. \rev{In the 80 Ridgecrest benchmark answers outside local retrieval, the default system named the $M7.1$ mainshock in only 44; the other 36 described the sequence without its largest event or refused (9), and one listed $M3.1$--$M4.4$ events among the ten largest.} Additionally, default prompts caused the model to hallucinate fluent but baseless physical mechanisms, such as falsely attributing Ridgecrest's seismicity to reservoir or fluid processes. \rev{Under the default, catalog-only condition, the average scores were generally poor: 1.35 for Qiaojia, 1.38 for Ridgecrest, and 0.78 for Maduo. Qiaojia and Ridgecrest struggled most with precise statistics (Category B at 0.70 and 0.70/3), while Maduo suffered from uniformly low scores across all categories (0.45--1.25/3). These baseline deficits, detailed in SM Table~S2, motivated the targeted prompt-engineering strategy in Section~\ref{sec:prompt}, which raised Category-D scores 1.6--2.2-fold and Maduo's overall average by +0.49.} Table~\ref{tabqa} gives six representative question/answer pairs \rev{from the seismology-prompt benchmark: four acceptable responses and the two residual error classes detailed in Section~\ref{sec:residual}}.

\begin{table*}
\caption{\rev{\textbf{Representative question/answer pairs from the seismology-prompt (post-fix) benchmark: four acceptable responses and the two residual error classes of Section~\ref{sec:residual}. Answers are quoted verbatim from the English-language post-fix answers, abridged (ellipses mark omissions); retrieval mode, condition, and question identifier in parentheses. The Outcome column gives the judge score (0--3; Section~\ref{sec:methods}) and the checked catalog fact. Full answers and scores: SM Table~S10 and the per-answer file in the Zenodo record.}}}
\label{tabqa}
\setlength{\tabcolsep}{3pt}
\begin{tabular*}{\textwidth}{@{\extracolsep{\fill}}p{0.26\textwidth}p{0.48\textwidth}>{\raggedright\arraybackslash}p{0.16\textwidth}@{}}
\hline
\textbf{Question} & \textbf{Answer (\rev{abridged, verbatim})} & \textbf{Outcome} \\
\hline
\rev{Which month contributed the highest maximum magnitude in the 2019 Ridgecrest sequence? (Global Search, catalog-only; B12)} & \rev{``The month of July 2019 contributed the highest maximum magnitude in the Ridgecrest earthquake sequence. The largest event recorded during this month was a significant earthquake with a magnitude of M7.10, which occurred on July 6, 2019. \ldots\ July 2019 also saw several other notable magnitudes, including M6.4 on July 4, 2019 \ldots''} & \rev{Correct, score 3 (catalog: $M7.1$, 6~July~2019 03:19~UTC; $M6.4$, 4~July)} \\
\rev{When did the closest events to the epicenter occur in the 2019 Ridgecrest sequence (spatial correlation)? (Local Search, catalog-only; C4)} & \rev{``The largest event in this sequence was recorded on July 6, 2019, with a magnitude of 7.10 \ldots\ The closest events to the epicenter of the largest earthquake (M7.10) occurred within the clusters identified on July 6, 2019 \ldots\ Largest Event: EVENT 2019-07-06T03:19:52.860 (M7.10)''} & \rev{Central claim correct, score 2: mainshock record and date exact; the named clusters and one event identifier are not in the catalog} \\
\rev{What is the general rupture/trigger mechanism of the sequence in the 2021 Qinghai Maduo $M_\mathrm{w}$~7.4 sequence? (Global Search, narrative-enriched; D7)} & \rev{``\ldots primarily associated with the KMPJ fault, which is recognized as the main fault responsible for the event. The earthquake occurred at a complex intersection of multiple fault systems, including the KMPJ fault, the Maduo-Gander fault, and the Jiuzhi fault. \ldots\ the aftershock sequence exhibited characteristics of bifurcation and bilateral rupture, particularly along the KMPJ fault.''} & \rev{Correct, score 3; grounded in the enrichment text \cite{Guan_2024} rather than pretrained priors} \\
\rev{Is the seismic activity in the 2019 Ridgecrest sequence associated with human industrial activities (e.g., hydraulic fracturing, wastewater injection, reservoir storage)? (Global Search, catalog-only; D2)} & \rev{``I am sorry but I am unable to answer this question given the provided data.''} & \rev{Correct refusal, score 3: the purely tectonic catalog contains no such evidence; default prompts had answered from background knowledge} \\
\rev{Which events are considered directly related to the largest magnitude event in the 2019 Ridgecrest sequence (e.g., aftershocks)? (Local Search, narrative-enriched; C10)} & \rev{``\ldots the largest recorded earthquake with a magnitude of Mw 7.1. This mainshock occurred on July 5, 2019, and was preceded by a notable Mw 6.4 earthquake on July 4, 2019. \ldots\ clusters of aftershocks were documented on July 6, 2019, and subsequent days \ldots''} & \rev{Wrong date, score 2: the $M7.1$ mainshock is on 6~July 03:19~UTC; 5~July is its local Pacific date (Section~\ref{sec:residual})} \\
\rev{When and where did the largest magnitude event in the 2021 Qinghai Maduo $M_\mathrm{w}$~7.4 sequence occur, and what was its magnitude? (Global Search, narrative-enriched; B2)} & \rev{``The largest magnitude event in the 2021 Qinghai Maduo earthquake sequence is the Maduo earthquake, which occurred on May 22, 2021. This significant seismic event registered a moment magnitude of 7.4.''} & \rev{Wrong, score 0: the catalog begins 1~June~2021 and its largest event is $M4.7$ (8~July~2021); the pre-catalog mainshock is presented as catalog data (Section~\ref{sec:residual})} \\
\hline
\end{tabular*}
\end{table*}

\section{Results: Narrative-Text Enrichment}
\label{sec:enrichment}

To assess the effect of narrative enrichment, each catalog was \rev{re-indexed with contextual literature (Baihetan Reservoir background for Qiaojia \cite{zhou_juxtaposition_2026}, Shelly \cite{shelly_highresolution_2020} for Ridgecrest, and Guan \emph{et al.} \cite{Guan_2024} for Maduo) and re-evaluated.} The effect is catalog-dependent and non-monotonic (Category-D scores in SM Table~S3\rev{; lenient convention, Section~\ref{sec:methods}}). Qiaojia's Category-D score rose \rev{(1.40 to 1.65)} but mainly reflects a more cautious tone matching the text's ``contested'' framing rather than genuine use of its specific named entities.
\rev{For Ridgecrest, the score fell (1.60 to 1.45).} When asked about industrial activity, the model simply refused to answer instead of using the newly provided text by Shelly \cite{shelly_highresolution_2020} to confidently deny the claim. While a blank refusal is the correct response when using only the \rev{raw catalog, since the bare catalog lacks mechanism data, it represents} a failure to utilize the enriched text, which contained the evidence needed to explicitly rule out the mechanism.
\rev{While Maduo's score increased (1.25 to 1.35)}, the enriched text introduced a major new mistake: the model confused the natural deep-fluid processes described by Guan \emph{et al.} \cite{Guan_2024} with human-induced hydraulic fracturing. This error did not occur when the model relied on the catalog alone. Additionally, under default settings, the model hallucinated the answer for the ``largest-magnitude event.'' It correctly named the real-world May 2021 mainshock, but because the provided catalog data did not start until 1~June, the model pulled this fact from its outside memory \rev{rather than from the data, a ``coincidentally correct'' error} that persisted even after prompt engineering.

\section{Results: Iterative Prompt Engineering}
\label{sec:prompt}

\rev{After identifying the failure modes in Sections~\ref{sec:capability} and~\ref{sec:enrichment}, we made four prompt and configuration changes, applied identically across all conditions. They target} magnitude fabrication via field confusion, historical-earthquake conflation, cross-catalog place-name contamination, and a scope-collapse bug in community detection. Each intervention was tailored to a diagnosed root cause; for instance, the scope-collapse issue was traced to a default GraphRAG clustering parameter that inadvertently discarded the majority of catalog entities. \rev{Diagnostic evidence, the details of each mechanism, and the exact prompt modifications are provided in SM Texts~S1--S4, and the complete configuration of each run is listed in SM Table~S9.}

\begin{figure}[t]
\centering
\includegraphics[width=.92\columnwidth]{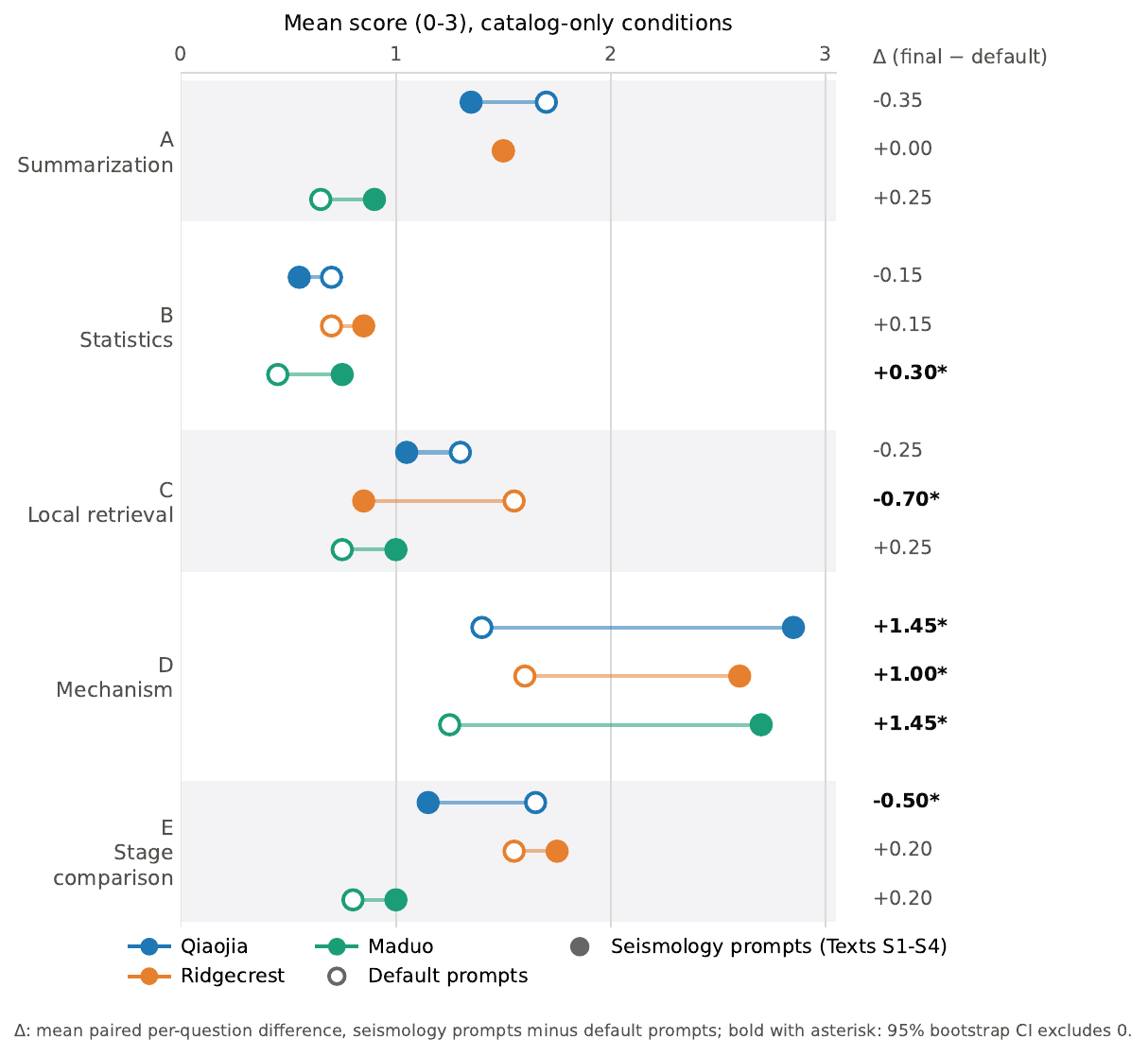}
\caption{\rev{\textbf{Per-category benchmark scores under default prompts (open circles) and after the Section~\ref{sec:prompt} prompt-engineering scheme (filled circles), catalog-only condition. Connectors join the before and after values; the right column ($\Delta$) gives the mean paired per-question change, final minus default, with an asterisk where the 95\% bootstrap interval excludes zero.}} \textbf{Physical-mechanism questions (Category D) improve most; precise-statistics questions (Category B) remain architecture-limited.}\label{fig04}}
\end{figure}

\rev{We re-indexed every affected condition and re-ran the full benchmark, scoring all answers as described in Section~\ref{sec:methods} (Fig.~\ref{fig04}; Table~\ref{tab05}).} All three targeted fabrication modes were \rev{eliminated}: \rev{no impossible magnitude value appeared in any of the 600 post-fix answers, versus repeated ``$M10.4$''/``$M12.8$'' before (the only out-of-range magnitudes quoted are real out-of-catalog events, Section~\ref{sec:residual}), no cross-catalog place-name contamination was found, and historical-earthquake conflation (previously $\sim$1 in 5 relevant answers) disappeared.} Despite the prompt fixes, two distinct types of hallucinations remained: presenting out-of-catalog events as real data, and inventing aggregate counts (detailed in Section~\ref{sec:residual}). \rev{Both the eliminated and the residual errors are behaviors of one model, \texttt{gpt-4o-mini}, under these prompts; a model with different pretrained knowledge may show different failure modes and respond differently to the same fixes \cite{Ji_2023}}. While scores for physical-mechanism questions (Category D) \rev{improved most, from 1.40 to 2.85 for Qiaojia and 1.60 to 2.60 for Ridgecrest, precise statistics (Category B) remained weak.} \rev{Category B scored between 0.55 and 1.15 out of 3, making it the lowest-performing category in all six test conditions.} The main issue is that the model severely undercounts totals and rates. This happens because a retrieval-then-synthesize system tries to answer math questions using a few text snippets rather than \rev{scanning the entire database, an architectural limitation rather than a prompt issue} \rev{(SM Text~S7.5)}.

The vector-RAG baseline (Table~\ref{tab05}) shows that \rev{GraphRAG's advantage lies where its community reports are designed to help: catalog-wide summarization} \rev{(Category~A: 1.35 vs.\ 1.15 on Qiaojia)} and temporal-stage comparison \rev{(Category~E: 1.75 vs.\ 1.00 on Ridgecrest)}, where top-$k$ chunk retrieval cannot synthesize catalog-wide structure and mostly returns hedged partial descriptions or refusals. Category~B is comparably weak for both architectures \rev{(0.55--0.85 across the Qiaojia and Ridgecrest catalog-only conditions; 0.75 vs.\ 1.00 on Maduo)}, confirming that this limitation is common to retrieval-then-synthesize systems rather than specific to GraphRAG. The baseline's \rev{near-perfect} Category-D scores (\rev{2.80--}3.00) reflect uniform honest refusals scored as correct because the catalogs contain no mechanism evidence. \rev{On Maduo the two systems are at parity (average 1.34 vs.\ 1.27), and overall GraphRAG and the baseline are statistically indistinguishable on all three catalogs (SM Table~S7)}. \rev{The baseline also reproduces the famous-mainshock intrusion of Section~\ref{sec:residual}, misdating Ridgecrest's $M7.1$ from pretrained knowledge while admitting that the retrieved chunks lack it, which places this error in the LLM rather than in GraphRAG.}

\begin{table*}
\caption{\textbf{Category-average scores after the targeted fixes (Section~\ref{sec:prompt}), all catalogs/conditions, and the vanilla vector-RAG baseline (catalog-only)\rev{; scoring procedure in Section~\ref{sec:methods}, per-question scores in SM Table~S10}.}}
\label{tab05}
\setlength{\tabcolsep}{3pt}
\begin{tabular*}{\textwidth}{@{\extracolsep{\fill}}lcccccc@{}}
\hline
\textbf{Condition} & \textbf{A} & \textbf{B} & \textbf{C} & \textbf{D} & \textbf{E} & \textbf{Avg} \\
\hline
Qiaojia catalog-only & \rev{1.35} & \rev{0.55} & \rev{1.05} & \rev{2.85} & \rev{1.15} & \rev{1.39} \\
Qiaojia +narrative & \rev{1.40} & \rev{0.55} & \rev{1.25} & \rev{1.70} & \rev{1.10} & \rev{1.20} \\
Ridgecrest catalog-only & \rev{1.50} & \rev{0.85} & \rev{0.85} & \rev{2.60} & \rev{1.75} & \rev{1.51} \\
Ridgecrest +narrative & \rev{1.75} & \rev{1.15} & \rev{1.40} & \rev{1.40} & \rev{1.50} & \rev{1.44} \\
Maduo catalog-only & \rev{0.90} & \rev{0.75} & \rev{1.00} & \rev{2.70} & \rev{1.00} & \rev{1.27} \\
Maduo +narrative & \rev{1.10} & \rev{0.55} & \rev{0.65} & \rev{1.70} & \rev{0.95} & \rev{0.99} \\
Qiaojia vector-RAG baseline & \rev{1.15} & \rev{0.85} & \rev{1.20} & \rev{3.00} & \rev{1.00} & \rev{1.44} \\
Ridgecrest vector-RAG baseline & \rev{1.45} & \rev{0.80} & \rev{1.20} & \rev{3.00} & \rev{1.00} & \rev{1.49} \\
Maduo vector-RAG baseline & \rev{1.00} & \rev{1.00} & \rev{0.90} & \rev{2.80} & \rev{1.00} & \rev{1.34} \\
\hline
\end{tabular*}
\par\vspace{1ex}\noindent
A: summarization; B: statistics; C: local retrieval; D: mechanism; E: stage comparison. Baseline rows share chunks, embeddings, LLM, and questions with the corresponding catalog-only condition (Section~\ref{sec:methods}); their Category-D scores reflect uniform honest refusals scored as correct. \rev{Under the stricter convention, which scores refusals on answerable questions as 0 rather than 1, the baseline averages fall to 0.84, 1.07, and 0.57, and the GraphRAG catalog-only averages to 1.13, 1.40, and 0.60 (both conventions for every condition: SM Table~S12).}
\end{table*}

\section{Residual Catalog-Specific Errors}
\label{sec:residual}

\rev{Per-answer scoring (SM Table S10) of the final pipeline (Section~\ref{sec:prompt}) leaves two residual error classes that the prompt fixes did not remove. Both are tied to the two catalogs with a globally famous mainshock, and both are concentrated in the narrative-enriched condition rather than in the catalogs as such; rates for every system and condition, with 95\% confidence intervals (CIs), are given in SM Table~S14 and examples in SM Text~S5. Neither class is an impossible value: no magnitude of 10 or more, the signature of the field-confusion failure, occurs in any of the 600 seismology-prompt answers, against 43 of the 600 default-prompt answers, and the out-of-range magnitudes that remain (e.g., $M_\mathrm{w}$~7.4) are real events outside the catalog.}

\rev{On Ridgecrest, the true $M7.1$ mainshock (6~July~2019, 03:19~UTC, the busiest day) is misidentified (placed on 4 or 5~July, or the 4~July $M6.4$ foreshock called the mainshock) in 48 of the 100 narrative-enriched answers (79\% of the 61 answers that name the mainshock; 95\% CI 67--87\%), but in only 4 of the 100 catalog-only answers (11\% of the 35 that name it). The default prompts show the same split (16 versus 1). The enrichment text \cite{shelly_highresolution_2020} dates events in UTC and never gives 5~July; the date the model supplies is the local Pacific date of the same event (5~July, 20:19 PDT). The error is therefore pretrained knowledge of a famous earthquake, expressed in local time, that surfaces once the background text makes the model narrate ``the mainshock'' instead of reading records, and that conflicts with the catalog's UTC timestamps. The correct answers in Section~\ref{sec:capability} and Table~\ref{tabqa} come from the catalog-only condition; in the enriched condition the same question type yields the wrong date in most answers.} 

\rev{On Maduo, whose catalog begins on 1~June~2021 with no May data, the real 22~May~2021 mainshock is presented as catalog data in 44 of 100 narrative-enriched answers (44\%, CI 35--54\%) but in only 1 catalog-only answer, and the enriching paper's total of 78,832 relocated events is quoted as the catalog's size in 24 narrative-enriched answers (the catalog holds 10,621); with default prompts the counts are 52 versus 4, and 50. Here the source is the ingested document itself, which states both figures: the model merges the scope of the paper with the scope of the catalog. The two classes thus share a trigger: narrative enrichment makes the model synthesize an account of the sequence rather than list records, and it fills that account from world knowledge (Ridgecrest) or from the enriching document (Maduo). The \texttt{HISTORICAL\_EVENT} fix of Section~\ref{sec:prompt} (SM Text~S4), which targets explicitly historical earthquakes, does not cover a mainshock that belongs to the sequence itself. The vector-RAG baseline reproduces the Ridgecrest misdating from pretrained knowledge in 1 of the 9 answers that name the mainshock (SM Text~S5.4) but fabricates far less (2--16\%), at the price of refusing 62--98\% of questions.} \rev{Extended examples are given in SM Text~S5.}

\section{Discussion and Conclusions}
\label{sec:conclusions}

This study demonstrates that GraphRAG can be applied directly, end-to-end, to earthquake catalogs that differ widely in scale, region, and origin, with no hand-designed schema, custom extraction rules, or per-catalog reconfiguration, to produce structurally complete knowledge graphs supporting flexible natural-language queries, with verified instances of precise record retrieval and grounded mechanism reasoning (Table~\ref{tabqa}), though answer reliability varies by catalog and question category (Table~\ref{tab05}). This is, to our knowledge, the first systematic demonstration of GraphRAG's schema-free construction capability transferring across independent earthquake catalogs \rev{with no ontology engineering}\rev{ and at a measured compute cost of 32--226 minutes of indexing and US\$0.7--4.8 of gpt-4o-mini API calls per catalog at list prices, with a median query latency of 44~s for Global Search and 7~s for Local Search (SM Table~S11)}, although downstream answer quality does not transfer uniformly.

Beyond capability, we show that reliability is not fixed once GraphRAG is applied out of the box, but can be measurably improved with modest, seismology-oriented prompt engineering. These findings point to a practical and affordable strategy for deploying LLM query tools on earthquake catalogs: start with the default GraphRAG pipeline, test it against a small ground-truth benchmark, and refine the prompts instead of building a new system from scratch. \rev{However, the gains are uneven. The approach removed the targeted hallucinations in all test cases, but the overall improvement survives the paired test for only one of the three catalogs. Because the original benchmark was used during prompt development, and the prompts were therefore tuned to its diagnoses, we also evaluated the finalized prompts on an unseen fourth catalog, the 2022 Luding $M$6.8 aftershock sequence (SM Text~S8; Table~S15). The revised prompts increased the score from 0.29 to 1.30/3 and substantially reduced the targeted errors, including impossible magnitudes and false references to an absent mainshock, although at the cost of more refusals. These results show that the prompt fixes generalize in suppressing the targeted failure modes, while the overall improvement depends on their prevalence in each catalog.} \rev{Narrative-text enrichment, by contrast, reduced post-fix overall scores for Qiaojia ($1.39\!\rightarrow\!1.20$/3), Ridgecrest ($1.51\!\rightarrow\!1.44$/3), and Maduo ($1.27\!\rightarrow\!0.99$/3), and it multiplied out-of-catalog mainshock intrusions on Maduo (44\% of enriched answers versus 1\% catalog-only; Section~\ref{sec:residual}), so it should be adopted only with per-catalog verification.}

\rev{Three limitations should be considered in practical deployment. First, precise statistical queries (Category B) remain unreliable even after prompt refinement, including on the independent Luding catalog (0.85/3; SM Table S15). Such calculations are therefore better handled by deterministic database or graph queries than by the LLM. Second, incorporating external literature can introduce information that is absent from, or inconsistent with, the catalog itself. As shown for Ridgecrest and Maduo in Section~\ref{sec:residual}, this can cause the model to present literature-derived information as catalog facts, whereas such errors are rare when only the catalog is provided. These results highlight the need to distinguish catalog-derived evidence from external contextual knowledge when using GraphRAG for seismological analysis. These limitations can be partly mitigated through several practical measures. Catalog time conventions should be explicitly specified, and external documents should be clearly scoped to prevent contextual information from being treated as catalog facts. Simple deterministic checks against the source catalog can provide an additional layer of validation: our post-processing screen detected 49\% of fabricated or out-of-scope answers (88\% for the Luding catalog) with 98\% precision. Third, our evaluation is limited to a single language model, \texttt{gpt-4o-mini}, which was used for indexing, querying, and the vector-RAG baseline (SM Table S9). The observed error rates and the effects of prompt refinement may therefore differ for other models. Nevertheless, the evaluation framework itself is model-independent and can be applied to other LLMs by validating their outputs against catalog-derived ground truth and systematically examining the remaining errors.}

\rev{Taken together, our results show that GraphRAG can automatically transform earthquake catalogs into queryable record-level representations, but its value lies primarily in answering open-ended questions rather than performing precise statistical analysis or recovering physical seismicity structure. Its overall performance is comparable to vector-RAG and varies across catalogs, while seismology-specific prompting improves reliability by reducing targeted failure modes. GraphRAG is therefore best used alongside, rather than in place of, deterministic catalog queries: the former supports flexible exploration and synthesis, while the latter provides reliable quantitative results.}

\section*{Data Availability}
All data and code used in this work, including the refined prompts, can be found at \url{https://doi.org/10.5281/zenodo.22906332}. \rev{The Supplementary Materials provide the complete GraphRAG configuration of every run, the 100 benchmark questions, the per-question score of every answer, the uncertainty analysis, per-category score tables under both refusal conventions, indexing time and cost, entity-type and community-size distributions, and the exact prompt-engineering modifications.}

\section*{Declaration of Competing Interests}
The authors declare that they have no conflicts of interest.

\section*{Acknowledgment}
Y. Zhou and S. M. Mousavi were supported by the Harvard Milton Fund. The authors thank the developers of the GraphRAG framework and the seismological data providers whose catalogs and publications made this study possible. The authors also thank the members of Plantcore.AI, whose insights during discussions helped inspire this work.

\bibliographystyle{IEEEtran}
\bibliography{references,references_extra}


\begin{IEEEbiography}[{\includegraphics[width=1in,height=1.25in,clip,keepaspectratio]{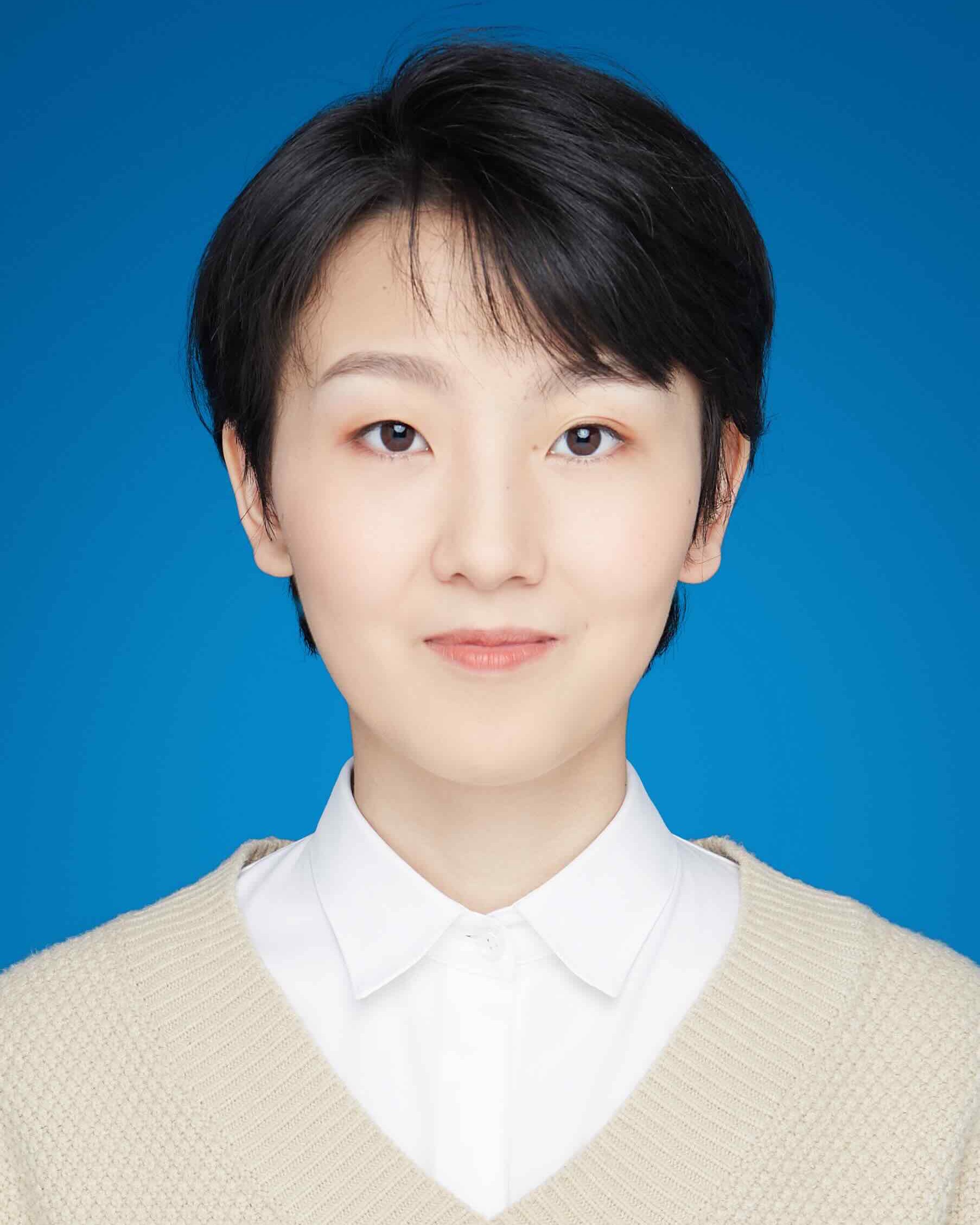}}]{Yuxin Zhou} is currently pursuing a doctoral degree with the State Key Laboratory of Earth System Numerical Modeling and Application, Department of Earth and Planetary Sciences, University of Chinese Academy of Sciences, Beijing, China, and is a visiting student with the Department of Earth and Planetary Sciences, Harvard University, Cambridge, MA, USA.

Yuxin Zhou's research interests include knowledge graphs, retrieval-augmented generation, large language models for seismology, and the automated analysis of high-resolution earthquake catalogs.
\end{IEEEbiography}

\begin{IEEEbiography}[{\includegraphics[width=1in,height=1.25in,clip,keepaspectratio]{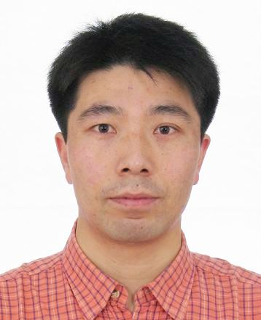}}]{Huai Zhang} is currently a Professor with the State Key Laboratory of Earth System Numerical Modeling and Application, Department of Earth and Planetary Sciences, University of Chinese Academy of Sciences, Beijing, China.

Prof. Zhang's research interests include large-scale numerical modeling of Earth systems, computational geodynamics, and earthquake physics. 
\end{IEEEbiography}

\begin{IEEEbiography}[{\includegraphics[width=1in,height=1.25in,clip,keepaspectratio]{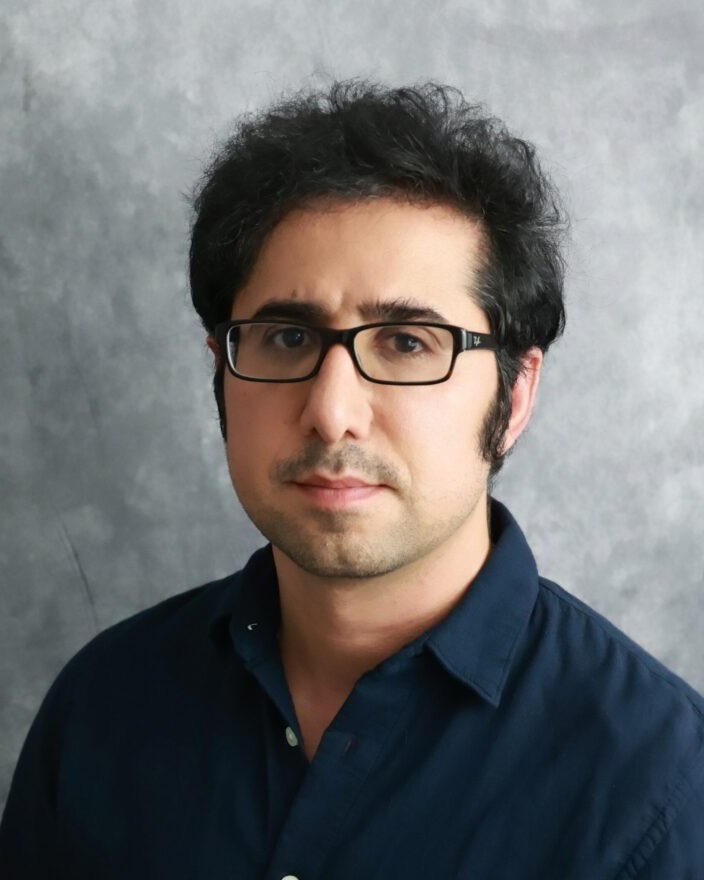}}]{S. Mostafa Mousavi} is currently an Assistant Professor with the Department of Earth and Planetary Sciences, Harvard University, Cambridge, MA, USA.

Dr. Mousavi's research is an interdisciplinary blend of seismology, statistics, and computer science, focused on extracting insights about Earth's physical processes from seismic signals. Dr. Mousavi uses artificial intelligence to analyze noisy, incomplete, high-dimensional data. 
\end{IEEEbiography}

\EOD

\end{document}